# $^{57}$Fe Mössbauer spectroscopy and magnetic measurements of oxygen deficient LaFeAsO

Israel Nowik and Israel Felner

Racah Institute of Physics, The Hebrew University, Jerusalem, 91904, Israel

and

V.P.S. Awana*, Arpita Vajpayee and H. Kishan

National Physical Labpratory, New Delhi-110012, India

**Abstract**

We report on the magnetic behavior of oxygen deficient $LaFeAsO_{1-x}$ (x~0.10) compound, prepared by one-step synthesis, which crystallizes in the tetragonal (S.G. P4/nmm) structure at room temperature. Resistivity measurements show a strong anomaly near 150 K, which is ascribed to the spin density wave (SDW) instability. On the other hand, dc magnetization data shows paramagnetic-like features down to 5 K, with an effective moment of 0.83 $\mu_B$/Fe. $^{57}$Fe Mössbauer studies (MS) have been performed at 95 and 200 K. The spectra at both temperatures are composed of two sub-spectra. At 200 K the major one (88%), is almost a singlet, and corresponds to those Fe nuclei, which have two oxygen ions in their close vicinity. The minor one, with a large quadrupole splitting, corresponds to Fe nuclei, which have vacancies in their immediate neighborhood. The spectrum at 95 K, exhibits a broadened magnetic split major (84%) sub-spectrum and a very small magnetic splitting in the minor subspectrum. The relative intensities of the subspectra facilitate in estimating the actual amount of oxygen vacancies in the compound to be 7.0(5)%, instead of the nominal $LaFeAsO_{0.90}$. These results, when compared with reported $^{57}$Fe MS of non-superconducting LaFeAsO and superconducting $LaFeAsO_{0.9}F_{0.1}$, confirm that the studied $LaFeAsO_{0.93}$ is a superconductivity-magnetism crossover compound of the newly discovered Fe based superconducting family.

*: Corresponding Author

Dr. V.P.S. Awana (Room 109), National Physical Laboratory, Dr. K.S. Krishnan marg, New Delhi-110012, India. Fax No. 0091-11-25626938: Phone no. 0091-11-25748709
e-mail-awana@mail.nplindia. ernet.in: www.freewebs.com/vpsawana/

## Introduction

Recent reports on superconductivity (SC) of up to 54 K in ReFeAsO (Re = La, Pr, Sm, Nd, Gd) have renewed the interest of the scientific community in the research of high $T_c$ superconductors, outside of the cuprates family [1-5]. Further, this discovery led the theoreticians to consider high $T_c$ systems outside the cuprates [6-7]. In the ReFeAsO system, SC is confined to the Fe-As layers and the charge carriers are provided from the doping of fluorine or alternatively by depletion of oxygen from the RE-O planes [1]. This is similar to high $T_c$ cuprate superconductors, where SC resides in $Cu\text{-}O_2$ planes and carriers are provided from other near by layers. One of the most striking properties of the new Fe based superconductors is their seemingly very high upper critical field ($H_{c2}$) [9,10]. In contrast to the cuprates, the SC Fe based compounds are prepared and studied by a limited number of research groups. This is due to the complicated synthesis route [1-5,9,10], which requires high pressure high temperature facilities (HPHT, 6 Gpa, 1350 $^0$C) [2,3] and two step normal pressure synthesis [1,4]. In fact the materials yet require optimization in their phase purity, and appearance of SC depends strongly on the F (Fluorine) content or alternatively on the oxygen deficiency. The Indian group, have managed to synthesize the nominal non-SC $LaFeAsO_{0.90}$ compound via a relatively simple one step route [11]. Here, we report the magnetization and $^{57}$Fe MS studies of this compound. The interpretation of the MS adopted the method used in our recent publication on the SC oxygen deficient $SmFeAsO_{0.85}$ sample, in which coexistence of SC and magnetic order in the Fe-As layers have been observed [10]. The Fe ions are very sensitive to their local environment, and exhibit two sub-spectra, their intensity defines the oxygen vacancies concentration (value of x).

## Experimental details

Stoichiometric amounts of better than 3 N purity of As, Fe, La metal and $La_2O_3$ were weighed and mixed thoroughly in nominal composition of $LaFeAsO_{0.9}$ ($Fe+As+0.3La_2O_3+0.4La$). A pellet powder, sealed in evacuated quartz tubes (better than $10^{-4}$ Torr), was heated at 500, 850 and 1100 $^0$C for 12, 12 and 33 hours respectively in a single step procedure [11]. The room temperature x-ray diffraction (XRD) pattern was taken on Rigaku mini-flex diffractometer. The resistivity measurements, in temperature range of 12 to 300 K, were carried out by a four-probe method in a close cycle refrigerator. Zero-Field-Cooled (ZFC) and Field-Cooled-Cooling (FCC) dc magnetic measurements in the range of 5-300 K were performed in a commercial (Quantum Design) super-conducting quantum interference device (SQUID) magnetometer. Mössbauer spectroscopy studies were performed by using a conventional constant acceleration drive and a 50 mCi $^{57}$Co:Rh source. The velocity calibration was done at RT with a α-Fe absorber and the isomer shifts (I.S.) values are relative to that of iron. The observed spectra were least square fitted by theoretical spectra, assuming a distribution of hyperfine interaction parameters, corresponding to in-equivalent iron locations differing in local environment.

## Results and Discussion

Figure 1 depicts the fitted and observed XRD patterns of the studied $LaFeAsO_{1-x}$ compound and its resistance versus temperature (R-T) is shown in the inset. The compound crystallizes in tetragonal structure (space group. P4/nmm) with lattice parameters: a = 4.0339(2) $A^0$ and c = 8.7346(7) $A^0$. The fitting parameters, and the sites positions of the various atoms, are given in Ref. 11. The R-T plot (Figure.1, inset) of $LaFeAsO_{1-x}$ shows the typical metallic behavior below 150 K and semiconductor like resistivity upturn below 100 K. The metallic step at 150 K is related to the spin density wave (SDW) ordered state of this compound [11]. Neutron diffraction (ND) studies indicate that below 150 K, the tetragonal structure becomes monoclinic and belongs to space group P112/n [12].

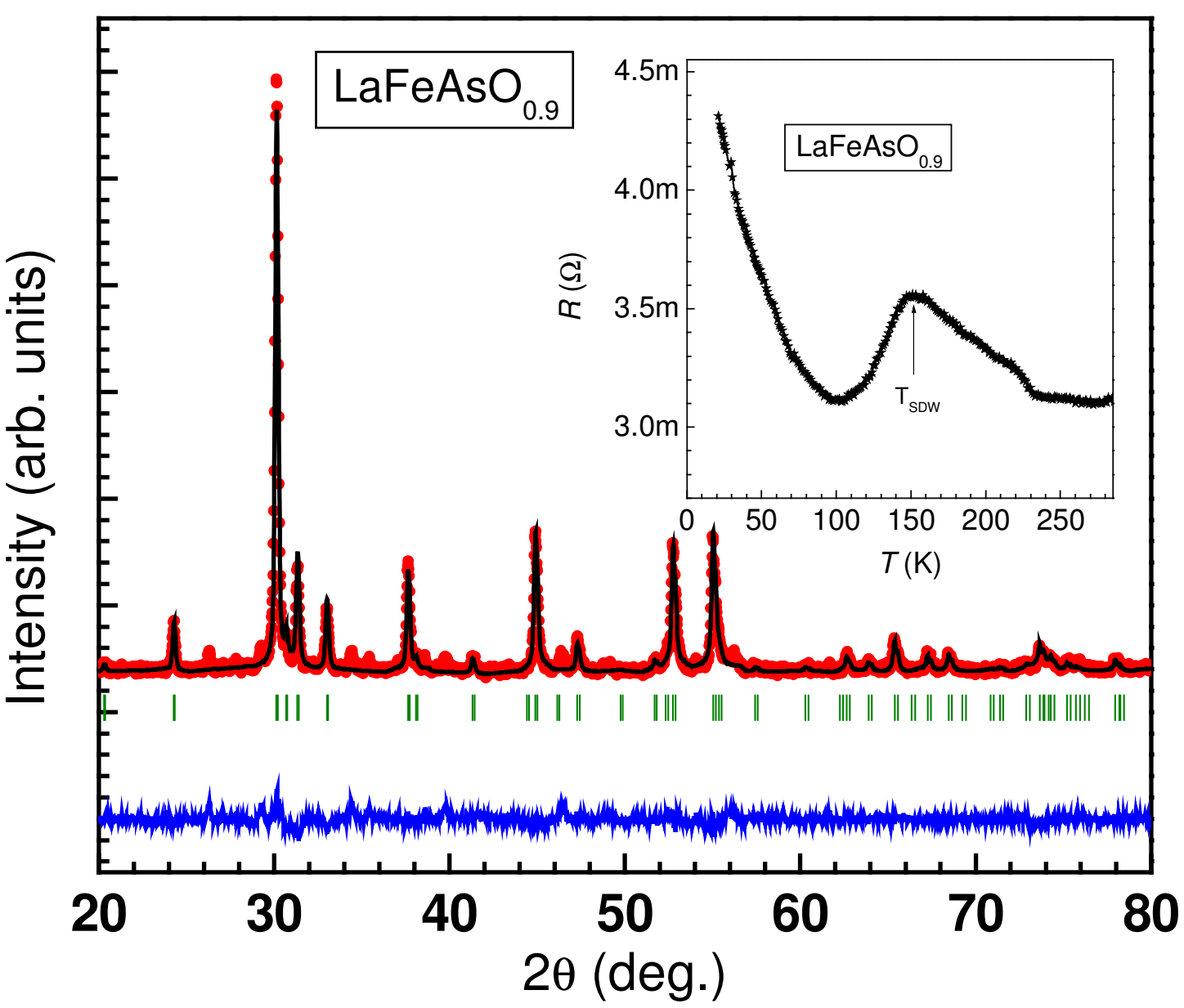

Figure 1. Fitted and observed X-ray diffraction patterns of nominal $LaFeAsO_{0.9}$, and its R-T plot (inset).

The temperature dependence of the magnetization for $LaFeAsO_{1-x}$ measured at 100 Oe, is shown in Fig. 2. The SDW features are not visible and the curve obtained has the typical paramagnetic (PM) shape and adheres closely to the Curie-Weiss (CW) law: $\chi(T) = \chi_0 + C/(T-\theta)$, where $\chi_0$ is the temperature independent part of $\chi$, C is the Curie constant, and θ is the CW temperature. The parameters obtained are: $\chi_{0=}$ 6.5*$10^{-6}$ emu/g*Oe, C=3.07*$10^{-4}$ emu*T/g*Oe which yields a PM effective moment of $P_{eff}$ =0.83 $\mu_B$ and θ=-2.8 K. Presumably, this effective moment corresponds to divalent Fe ions since all other ions are non-magnetic. This $P_{eff}$ is consistent with the low ordered moment (~0.25 $\mu_B$) at the Fe site in LaFeAsO deduced from ND, μSR and Mossbauer studies. [12-13]

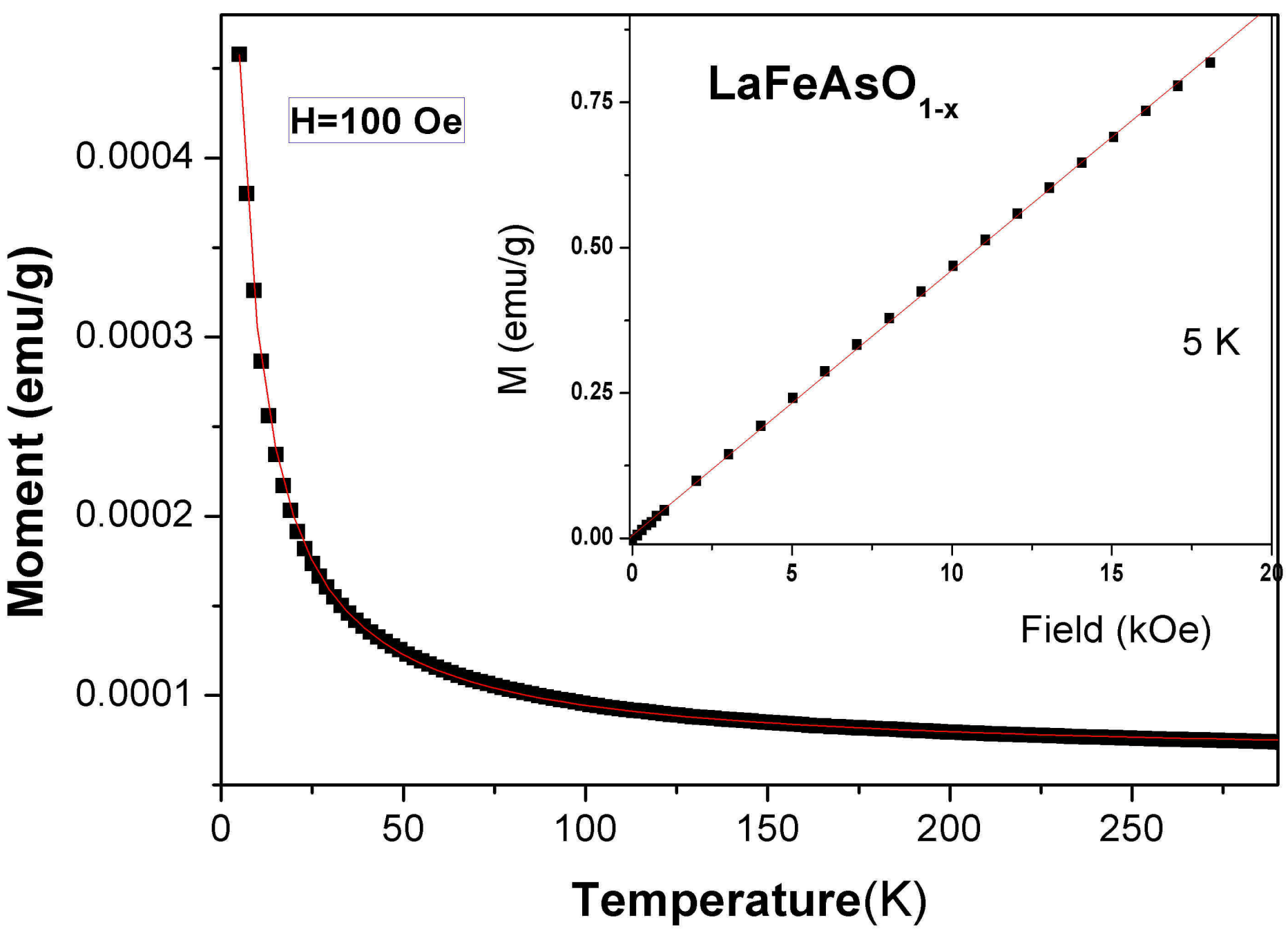

Figure.2. The temperature dependence of the magnetization for $LaFeAsO_{1-x}$ and the isothermal M(H) measured at 5 K(inset).

The MS spectra of $LaFeAsO_{1-x}$, measured below and above the magnetic transition are shown in Figure 3, and the deduced hyperfine interaction parameters are listed in Table 1. In both tetragonal P4/nmm and monoclinic P112/n structures each Fe, which resides in the 2b or 2f crystallographic positions respectively, has two oxygen ions as nearest neighbors along the crystal c-axis with the shortest Fe-O distance of c/2~4.2035 $A^0$. Each spectrum in Figure 3 is composed of two sub-spectra and the isomer shift (I.S.) values for both sites are typical to a divalent low-spin state for all Fe ions. At 200 K the intense sub-spectrum (~88%) is almost a singlet, and corresponds to those iron nuclei, which have two oxygen ions in their close vicinity. The second quadrupole doublet corresponds to iron nuclei, which have vacancies in their immediate neighborhood. In the case of random distribution of vacancies, the probability of an iron ion (in this layered structure) to have one or two vacancies as first nearest neighbor is 2x(1-x) and $x^2$, respectively. Thus the intensity of the minor site in the 200 K spectrum [12.0(3)%] indicates that in this material x=0.065 (and not 0.10). The probability to have iron with two neighboring vacancies is small and negligible. At 95 K both sites are magnetically ordered and the hyperfine parameters deduced are presented in Table 1. The major sub-spectrum was analyzed in terms of a very asymmetric Gaussian distribution of magnetic hyperfine fields ($H_{eff}$). The field distribution results probably due to a partly incommensurate spin density wave [14] caused by the vacancies. The value of $H_{eff}$ at maximum probability is ~50 kOe, similar to the low fields observed for the non-SC and magnetic LaFeAsO compound [15]. The minor subspectrum [16(2)%] has a large quadrupole splitting, like at 200 K, with a small magnetic splitting, see Table 1.

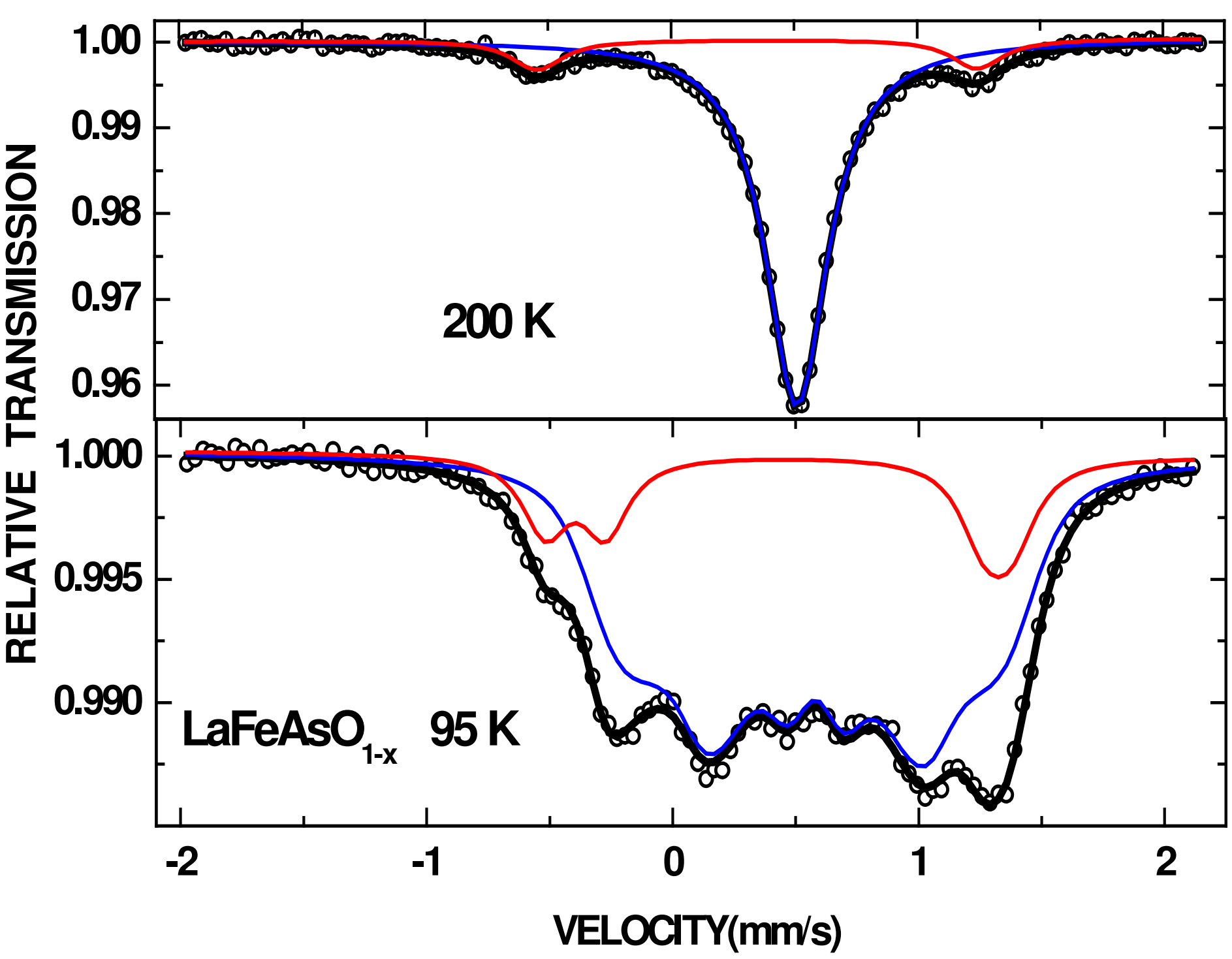

Figure 3. The Mössbauer spectra of $LaFeAsO_{1-x}$ measured at 95 and 200 K.

Table 1. The hyperfine interaction parameters for $LaFeAsO_{1-x}$ derived from the analysis of the Mössbauer spectra. INT, GAM, I.S., EQ and $H_{eff}$ stand for relative intensity, line width, isomer shift, quadrupole parameter (EQ=eqQ/4) and magnetic hyperfine field.

| T ( K) | INT(1) (%) | GAM mm/s | I.S. (1) mm/s | EQ(1) mm/s | $H_{eff}$ (1) (kOe) | INT(2) (%) | GAM mm/s | I.S.(2) mm/s | EQ(2) mm/s | $H_{eff}$ (2) (kOe) |
|---|---|---|---|---|---|---|---|---|---|---|
| 95 | 84 | 0.25(1) | 0.57(1) | 0 | 51(2) | 16(2) | 0.25(1) | 0.46(1) | -0.86(1) | 8 (1) |
| 200 | 88 | 0.30(1) | 0.51(1) | 0 | 0 | 12.0(3) | 0.25(1) | 0.35(1) | 0.89(1) | 0 |

The hyperfine parameters of the major sub-spectrum are identical at both temperatures and are very similar to values obtained in other SC compounds such as $SmFeAsO_{0.85}$ [10] $LaFeAsO_{0.89}F_{0.11}$ [15] and LaFePO [16]. In these samples the MS exhibit only one singlet, since no vacancies exist and therefore no second sub-spectrum is achieved. The vacancies obtained in our sample induce two types of Fe ions, and thus permits a direct insight into the various Fe ions behavior. Similar two sub-spectra were obtained in the oxygen deficient $SmFeAsO_{0.85}$ [10].

In conclusion we present a detailed study of the magnetic nature of oxygen deficient $LaFeAsO_{1-x}$ compound, by means of resistivity, dc magnetic measurements and Mossbauer spectroscopy studies. The resistivity curve shows a peak at 150 K, which is related to the SDW instability, exists in this sample. On the other hand, no anomaly is observed in the dc M(T) plot which rather shows a PM –type behavior in the entire temperature range. Due to oxygen vacancies, the MS spectra below and above 150 K are composed of two sup-spectra from which the oxygen vacancies concentration can be deduced. It appears that x=0.065 instead of the nominal composition of x=0.10. The isomer shift values for both sites are typical to a divalent Fe in the low-spin state. At 95 K, both Fe sites are magnetically ordered and the major sub-spectrum was analyzed in terms of an asymmetric Gaussian distribution of magnetic hyperfine fields, with the maximum probability of $H_{eff}$ ~50 kOe.

**Acknowledgments**: This research is partially supported by the Israel Science Foundation (ISF, 2004 grant number: 618/04) and by the Klachky Foundation for Superconductivity. Authors from NPL, India thank their director Professor Vikram Kumar for encouragement.